\def\year{2015}
\documentclass[letterpaper]{article}
\usepackage{arabtex}
\usepackage{utf8}
\usepackage{aaai}
\usepackage{times}
\usepackage{helvet}
\usepackage{courier}
\usepackage{color, colortbl}
\usepackage{array}
\usepackage{multirow}
\definecolor{Gray}{gray}{0.9}
\definecolor{White}{gray}{1.0}
\usepackage{url}
\usepackage[hide]{chato-notes} 
\usepackage{graphicx}

\begin{document}
%
\setcode{utf8}
\title{\#FailedRevolutions: Using Twitter to Study the Antecedents of ISIS Support}

\author{
Walid Magdy, Kareem Darwish, and Ingmar Weber\\
Qatar Computing Research Institute\\
Qatar Foundation, Doha, Qatar\\
{wmagdy,kdarwish,iweber}@qf.org.qa
}
\maketitle
\begin{abstract}
Within a fairly short amount of time, the Islamic State of Iraq and Syria (ISIS) has managed to put large swaths of land in Syria and Iraq under their control. To many observers, the sheer speed at which this ``state'' was established was dumbfounding. 
To better understand the roots of this organization and its supporters we present a study using data from Twitter. 
We start by collecting large amounts of Arabic tweets referring to ISIS and classify them into pro-ISIS and anti-ISIS. This classification turns out to be easily done simply using the name variants used to refer to the organization: the full name and the description as ``state'' is associated with support, whereas abbreviations usually indicate opposition. We then ``go back in time'' by analyzing the historic timelines of both users supporting and opposing and look at their pre-ISIS period to gain insights into the antecedents of support. To achieve this, we build a classifier using pre-ISIS data to ``predict'', in retrospect, who will support or oppose the group. 
The key story that emerges is one of frustration with failed Arab Spring revolutions. ISIS supporters largely differ from ISIS opposition in that they refer a lot more to Arab Spring uprisings that failed. We also find temporal patterns in the support and opposition which seems to be linked to major news, such as reported territorial gains, reports on gruesome acts of violence, and reports on airstrikes and foreign intervention.
\end{abstract}

\section{Introduction}

The recent rise and territorial gains of the ``Islamic State of Iraq and Syria'' (ISIS) (or simply the ``Islamic State'') has sparked significant interest in the group.  One particular aspect of interest in ISIS is related to the profile of individuals who are likely to join or merely support them.  Since the events of 9-11, there has been wide interest in identifying necessary and sufficient traits individuals may posses that are likely to propel them to support or join violent militant organizations. Contrary to popular belief, social psychology studies and reviews suggest that individuals who end up joining such organizations are typically more educated, financially better off, generally more accomplished than average, and more exposed to Western culture~\cite{Louis2009}.  Many studies have looked at whether such individuals suffer from psychological disorders, but found no evidence of such~\cite{horgan2003search}.  On the contrary, they generally exhibit higher than average psychological strength and are far less traumatized by incarceration, interrogation, or imprisonment~\cite{miller2006terrorist}.  In other words, they are not just normal people, but rather they are normal people with better than average fortunes.  Since millions of people fit this description and the overwhelming majority of them do not resort to violence, psychological traits are insufficient to identify people with violent tendencies. In a study of a leftist group in India which resorted to violence~\cite{Sarangi2005}, the twelve members of the group who were interviewed provided similar personal narratives in which they described structural deficiencies in their societies including corruption, oppression, and lack of empowerment for certain segments of the society.  Consequently, their personal narratives painted a picture of those engaged in violence as ``heroes'' who are confronting these deficiencies.  The findings of Sarangi and Alison~\shortcite{Sarangi2005} are in line with the findings of many other studies that show that people who resort to violence do so to eliminate what they perceive as injustices~\cite{PlousZ05}.  Along the same lines, there has been a recent story that was publicized in the media concerning a man called Ahmed Al-Darawy, a successful and affluent 38 year-old former manager in a multinational company in Egypt, who was killed while fighting alongside ISIS\footnote{\scriptsize \url{http://www.ft.com/cms/s/2/97130d46-7952-11e4-9567-00144feabdc0.html}}.  Al-Darawy was a former police officer, who left the service to join the revolutionaries who toppled the former Egyptian president Hosni Mubarak, and he later ran for elected office in an embrace of democracy.  Interviews with his friends after his death painted a picture of a person who was disillusioned and angry, particularly after the military intervention against the democratically elected president, Mohamed Morsi. Al-Darawy is highly prototypical of people who resort to violence.  Juergen Todenhoefer, a German journalist who recently spent 10 days with ISIS, provided confirming testimony in which he said ``These [ISIS fighters] are not stupid people. One of the people we met had just finished his law degree, he had great job offers, but he turned them down to go and fight.\footnote{\scriptsize \url{http://www.newsweek.com/german-journalist-returns-time-isis-chilling-stories-293781}}'' There is evidence in the literature that support for such militant groups is mostly political, rather than religious~\cite{Abdulla07}. 
In short, Louis~\shortcite{Louis2009} attributes the motivation to pursue violent means to the following beliefs: 1) ``Alternatives to terror do not work''; and 2) ``Terrorism can achieve social change''.  

In this paper, given Arab Twitter users (tweeps) with explicitly expressed positions supporting or opposing ISIS, we seek to identify distinguishing features from their Twitter profiles that foretell their future positions prior to authoring the first tweet overtly declaring their stance.  We examine the interests of both groups before and after they started explicitly supporting or opposing ISIS. In line with the findings in the literature, we attempt to analyze the motivations that prompt people to support ISIS (not necessarily join them), namely identifying perceived injustices that act as triggers.  

We conducted the study on approximately 57 thousand tweeps, who authored nearly 123 million tweets. The contributions of this paper are as follows:
\begin{itemize}
\item We determine distinguishing language that signals current support or opposition for ISIS.
\item We train a classifier that can predict future support or opposition of ISIS with \%87 accuracy.
\item We identify potential reasons for subsequent support of ISIS and the sources of opposition to the group.
\item We attempt to account for the profiles of people who currently support or oppose ISIS.
\item We illustrate some of the underlying phenomena in the group's social media engagement.
\end{itemize}

Our analysis is broken down into two parts. After discussing related work and describing our data collection, we first present a \emph{global trend analysis} of trends in pro- vs.\ anti-ISIS tweets. This provides insights into the effect of external events, such as reports of territorial gains or videos of beheadings, on the mobilization of supporters. After this, we present an \emph{individual historic analysis} where we put particular emphasis on the pre-ISIS period.

\section{Background}
\subsection{ISIS and the Syrian Conflict Online}
\begin{figure}
\centering
\includegraphics[width=0.7 \columnwidth]{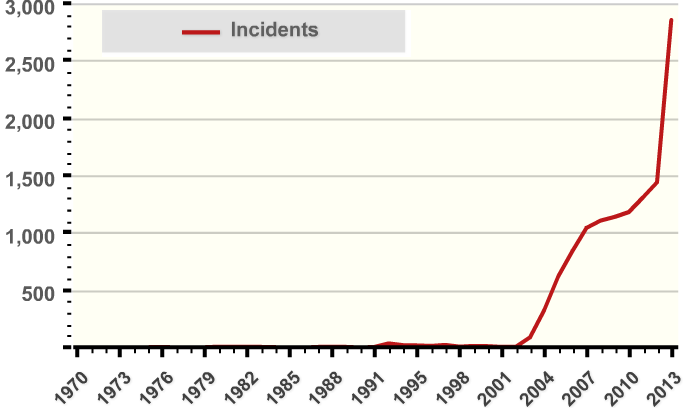}
\caption{\label{iraqTerror}Number of terrorist attacks in Iraq per year (source: Global Terrorism Database\footnote{\scriptsize \url{http://www.start.umd.edu/gtd/}})}
\end{figure}

Militant groups have long used the internet and social media for communication, information gathering, recruitment, and social network engagement~\cite{Cohen-Almagor12}. Though, specific literature on the use by ISIS of social media is scant, there are strong indicators that the group seems to be fairly media savvy with strong presence on different social media platforms\cite{shane2014isis}.  For example, ISIS (or its affiliates) maintain many Twitter accounts that propagate the group's message in several languages.  When accounts are taken down, new ones are quickly created to replace them\cite{shane2014isis}. ISIS also produces high quality videos in several languages as a recruitment tool.  For example, the group produced an English video entitled ``There is No Life Without Jihad'' that featured Western fighters that behoove others to join them\cite{kohlmann2014profiles}.  Much of the propaganda of ISIS is focused on reporting on military successes and ISIS's success in implementing different aspect of Sharia (Islamic law). Though they were not the first to engage in this kind of media campaigns, they have eclipsed other militant groups, such as Al-Qa'ida, in attracting new members.  Their recruitment success can be mostly attributed to their military victories, particularly the capture of Mosul, Iraq's second largest city\cite{kohlmann2014profiles}. Kohlmann and Alkhouri\shortcite{kohlmann2014profiles} provide accounts of several individuals who either joined or attempted to join ISIS or other militant groups in Iraq and Syria.  In most accounts, individuals expressed strong anti-American sentiment including one who was filmed tearing his American passport and burning it, another who said that killing American soldiers was justified, and yet another who spoke of a war that the West was waging against Muslims. Disdain for US intervention in the Middle East seems to be a common theme among Arabs in specific and Muslims in general\cite{jamal2014anti}. Further, US military intervention in Iraq and Afghanistan seems to coincide with a sharp increase in terror attacks in both countries.  Figure~\ref{iraqTerror} shows a steady rise in terrorist attacks in Iraq starting in 2003 (the year that the US invaded Iraq).
In most cases, individuals joining ISIS obtained information from the internet which came in the form of tweets from notable individuals, personal contact over social media, and online videos and lectures.  Thus, there are indications that the media strategy of ISIS is effective in attracting supporters~\cite{kohlmann2014profiles}.  The effectiveness of the social media strategy of ISIS and the boost it received due to the US bombing campaign against it has been reported on in the news\footnote{\scriptsize \url{http://www.reuters.com/article/idUSL1N0RI0R520140917}}.

Though not explicitly focused on ISIS, other studies have looked at Twitter communication in Syria during the ongoing conflict. O'Callaghan et al.~\shortcite{OCallaghanPGCCC14} performed a network analysis of 652 Twitter accounts of Syrians engaged in the conflict. These accounts were identified using Twitter lists and their analysis revealed four general communities, namely: (i) pro-Assad, (ii) Kurdish, (iii) secular/moderate opposition, and (iv) Islamists, including supporters of ISIS. Their analysis, performed before the major territorial gains of ISIS, also involved content analysis of the Freebase\footnote{\scriptsize \url{https://developers.google.com/youtube/v3/guides/searching_by_topic}} concepts assigned to YouTube videos that were uploaded by the Twitter users. Our analysis takes a different angle and tries to identify a wider range of users (57k, extracted from an initial 165k) and, importantly, tries to \emph{go back in time} to understand the antecedents of joining or merely supporting a particular group. Another study has investigated the reactions on Twitter in response to the 2013 Sarin Gas Attacks near Damascus \cite{TyshchukWLJK14}. Among other things, they observed that ``there was no immediate polarization of opinions following the event'', and that ``Twitter communities were too sparse to produce substantial amount of social pressure to force an opinion/opinion shift''. Finally, Morstatter et al.\ \cite{MorstatterPLC13} have used geo-tagged tweets from Syria to evaluate the bias between different sources and sampling methods of Twitter data.

\subsection{Using Online Data to Study Political Polarization}
Our pro- vs.\ anti-ISIS angle can also be seen as a study of online polarization. Polarization, usually along the political left vs.\ right dimension, has a long research history. One quantitative approach is the NOMINATE score \cite{poole1985spatial}. It uses roll call data voting history from the U.S.\ senate and house of representatives to can quantify the statement that U.S.\ politics follows a 1-dimensional left-to-right schema. Focusing on online data, polarization in U.S.\ politics has been studied in the setting of blogger networks \cite{Adamic2005} and Twitter \cite{ConoverRFGMF11} to name two influential studies. Geographically and topically closer to ISIS, online polarization and political violence has been studied using Twitter data from Egypt \cite{WeberGB13,Borge-HolthoeferMDW15}. The conflict between Israel and Palestine and how it plays out on Twitter has also recently been studied \cite{ZheIngmar14}. 
Though these studies describe differences in the users who are part of the two camps, none of these studies look at the \emph{antecedents} or attempts to explain what \emph{makes} a user join or support one side rather than another.

\section{Data Collection}

\begin{figure}
\centering
\includegraphics[width=0.6 \columnwidth]{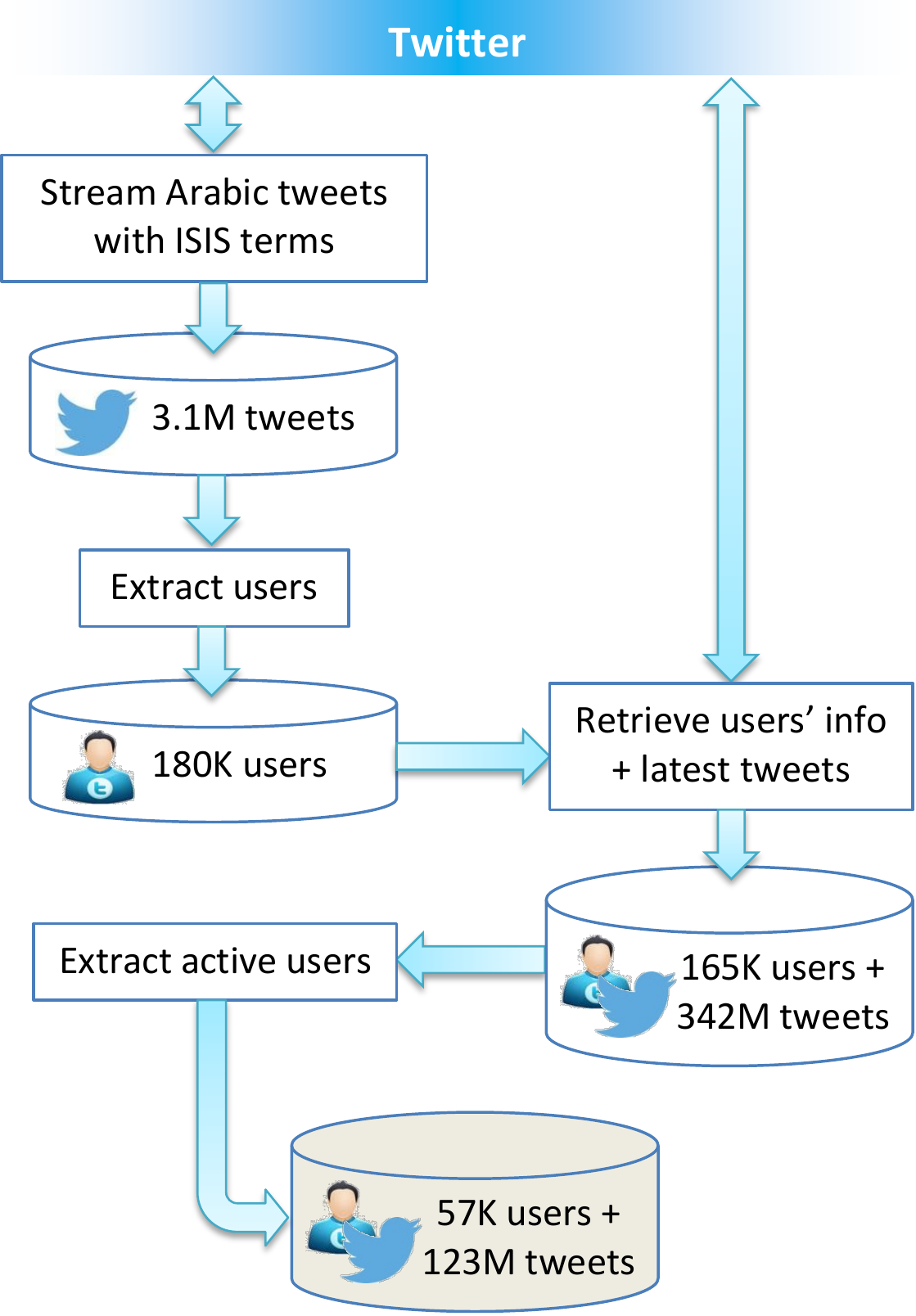}
\caption{\label{DataCollection}Steps performed for data collection}
\end{figure}

Since the majority of ISIS supporters come from the Arab World, we focused our study on Arabic Twitter profiles. Figure~\ref{DataCollection} illustrates the steps we followed for data collection. We first start by identifying tweets in Arabic by continuously searching the Twitter Rest API with the query ``lang:ar".\footnote{\scriptsize Using the language operator ``lang:'' on Twitter's Streaming API would have resulted in a dramatically reduced data set as the operator is provided on a post-sample basis, i.e., \emph{after} sampling down to 1\% of tweets.} From this set of Arabic tweets, we then collected all tweets mentioning ISIS by any of its name forms between mid October, 2014 and the end of December, 2014.  The name variations were of two types, namely: those that used the full name of the group such as ``\<الدولة الإسلامية>'' (Aldawla Alislamiya -- ``Islamic State'') and ``\<الدولة الإسلامية في العراق والشام>'' (Aldawla Alislamiya fi Aliraq walsham -- ``Islamic State in Iraq and the Levant''), and those that used the abbreviated version of the name such as ``\<داعش>'' (da'esh -- Arabic acronym for the group), , ``\<داعشي>'' (da'eshy -- from da'esh), and ``\<دواعش>'' (dawa'esh -- plural of da'eshy).

%
In all, we collected 3.1 million Arabic tweets mentioning ISIS in some form, authored by more than 250k users between Oct.\ 13, 2014 and Dec. 31, 2014. This data underlies our tweet-centric analysis. From these tweets we identified 180k tweeps during the period October 13, 2014 and November 1, 2014. Of the 180k tweeps, only 165k had active accounts for which we could obtain historic tweets through the Twitter Rest API. By Nov.\ 17, 2014, the remaining 15k accounts were either deleted by the user, suspended by Twitter, or had their privacy settings set changed to ``protected''. 

Next, we were interested in distinguishing between those who support and those who oppose ISIS. We found that the name used to refer to the group was telling of whether a person supports or opposes them.  A native Arabic speaker, who is well acquainted with the topic, judged a random sample of 1,000 tweets that was spread uniformly across all days for each condition as positive, negative, or neutral. Positive tweets included ones that praise the group, spreads their messages, or report on them positively.  Negative tweets include ones that attack the group and spreads negative news about them. Neutral ones included those that merely report news or spam tweets that use ISIS related hashtags to increase their exposure.

Table~\ref{searchterms} presents the percentage of tweets that support or oppose ISIS based on whether they use; a) the full name (or its variants); or b) the abbreviated name (or its variants).  As shown, using the full name of group is a strong indicator of support for ISIS (93\%), and using the acronyms is a general indication of opposition (77\%). For the 23\% of tweets that use the acronym, 7.5\% support ISIS and 15\% do not reveal any specific leaning.  The fact that both pro- and anti-ISIS tweeps carefully choose how they refer to ISIS has been observed by others before\footnote{\scriptsize See e.g. 
\url{http://www.ibtimes.com/isil-isis-islamic-state-daesh-whats-difference-1693495},
\url{http://www.independent.co.uk/9750629.html},
 and \url{http://cnn.com/2014/09/09/world/meast/isis-isil-islamic-state/}}. 
Based on the observation that using the full name mostly indicates support and the abbreviated name generally indicates opposition, out of the 3.1 million tweets that we collected, 1.2 million and 1.8 million tweets were supporting and opposing ISIS respectively.  For all tweeps, we scraped the last 3,200 tweets that they authored, totaling 324 million tweets, along with other Twitter profile information such as profile creation date and user declared location.

From the full set of tweeps, we retained 56,960 who: 1) authored 10 or more tweets (out of 3,200 tweets) that mention ISIS, and 2) strictly used either the long name or strictly used the abbreviated name of ISIS in at least 70\% of their tweets that mention the group.  

These two conditions were meant to ensure that we perform our study on tweeps who are both engaged in the topic and have voiced clear positions towards ISIS.  To validate the definiteness of user positions (pro- or anti-ISIS) based on the two conditions, we had the same judge automatically label 50 random tweeps from each group.  The judge labeled tweeps as pro- or anti-ISIS based on the tweets that they authored mentioning the group. We found that \emph{all} 100 tweeps were correctly labeled, which suggests that the conditions yield high labeling precision.  Through this, we automatically labeled 11,332 users as pro-ISIS and 45,628 users as anti-ISIS.  The tweeps meeting these conditions authored 123 million (out 324 million we collected).  

\begin{table}
\begin{center}
\begin{tabular}{lccc}
\hline
Search query & Anti-ISIS & Pro-ISIS & Neutral/Spam \\\hline
``ISIS'' & 77.3\% & 7.5\% & 15.2\% \\
``Islamic State'' & 1.2\% & 93.1\% & 5.7\% \\\hline
 \end{tabular}
\end{center}
\caption{\label{searchterms}Percentage of pro-/anti-ISIS tweets given the presence of either an abbreviated form (``ISIS'') or the full name (``Islamic State'')}
\end{table}

\section{Pro/Anti ISIS Tweet Counts and Temporal Trends}
\begin{figure*}[ht]
\centering
\includegraphics[width=1.0\textwidth]{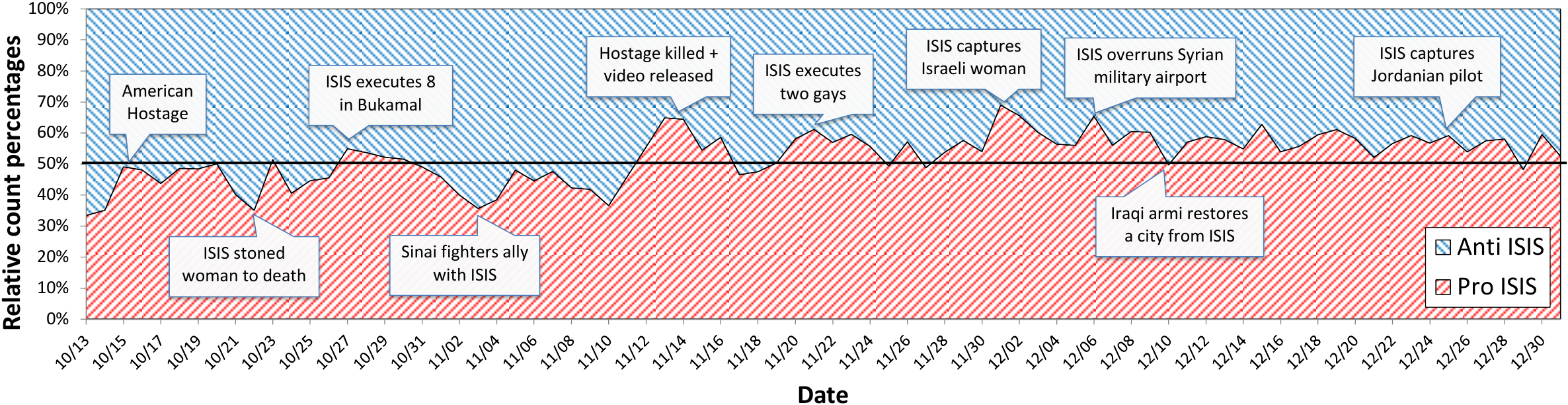}
\caption{\label{TweetCounts}Pro/anti ISIS tweet counts}

\includegraphics[width=1.0\textwidth]{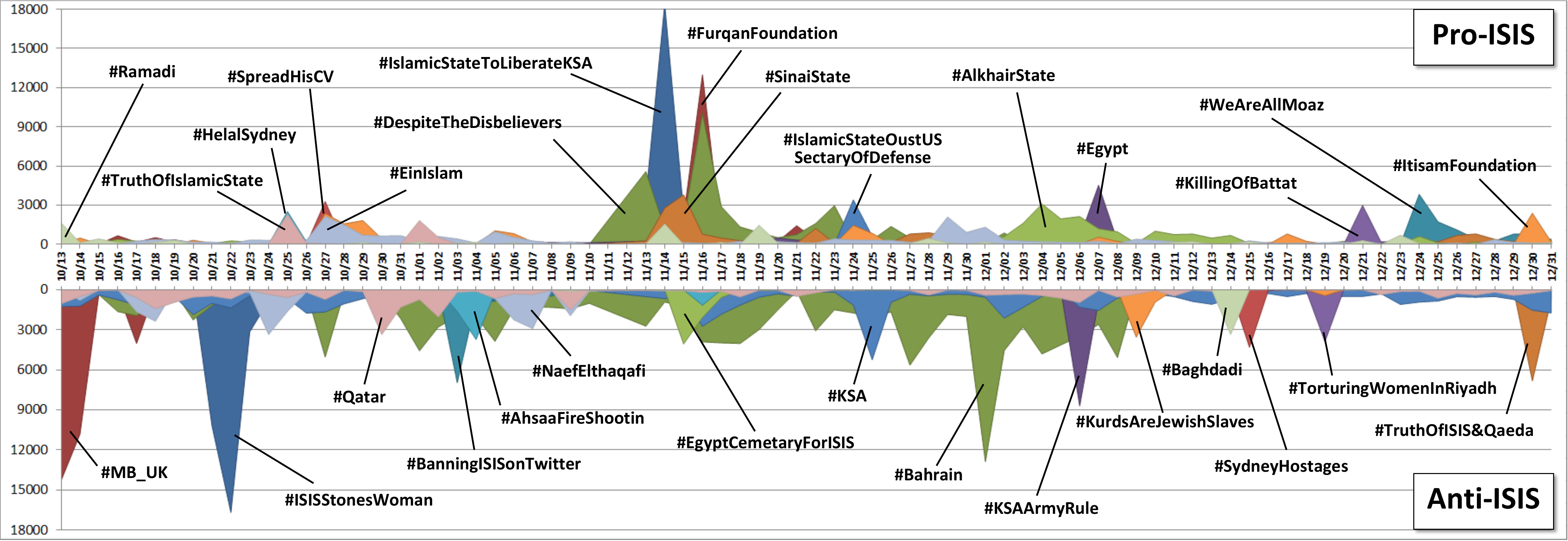}
\caption{\label{HashtagTimeline}Top hashtags used by pro- and anti- ISIS tweeps with their counts}
\end{figure*}

In this section, we explore the temporal tweeting trends for those who support or oppose ISIS. We attempt to correlate these trends with events that happened during the period of study.  Figure~\ref{TweetCounts} shows the tweet counts for both groups from Oct.\ 13, 2014 to Dec.\ 31, 2014. 

Figure~\ref{TweetCounts} shows the relative percentages of pro- and anti-ISIS tweets. The counts for both groups are weakly correlated with a Pearson's \emph{r} of 0.27.  Pearson's \emph{r} ranges between -1 (complete negative correlation) and 1 (complete positive correlation) and 0 implies no correlation.  As can be seen, the fraction of pro-ISIS tweets steadily increased over the time period, while anti-ISIS tweets has steadily decreased. However, it is not clear if the increase in pro-ISIS tweets is organic or not.  Figure~\ref{TweetCounts} lists some of the major news stories that appeared on the days when there were peaks for ISIS supporters and opposers.  As can be seen from the events, anti-ISIS tweets generally peaked when news of ISIS human rights violations emerged such as the killing of hostage, accounts of torture, or reports of the enslavement of Yazidi women.  On the other hand, pro-ISIS tweets generally peaked in conjunction with the release of propaganda videos and major military achievements.  Though some connection between events on the ground and shifts in the relative popularity can be observed, it is not clear how strong it is, and it is very hard to determine causality.

Figure~\ref{HashtagTimeline} shows the top 15 peaking hashtags in association with pro- and anti-ISIS tweets, along with their frequency counts. To obtain the top hashtags, we: a) extracted the hashtags from the original set of 3.1 million tweets; b) eliminated hashtags that are name variants of ISIS; c) we obtained and sorted hashtag-per-day counts; and d) we retained the top 15 that had the highest hashtag-per-day counts (and not the global count over the entire period).  In other words, if a hashtag was used 1,000 times in one day, it would be preferred over another that was used about 100 times everyday for 20 days.  

The upper part of Figure~\ref{HashtagTimeline} shows the top 15 hashtags (translated from Arabic) used by pro-ISIS tweeps cover a variety of topics such as (each hashtag is followed by its rank): threats and claims of victory against opponents as in \#ISIS\_to\_liberate\_KSA (1), \#ISIS\_ousts\_US\_sectary\_of\_defense (7), and \#killing\_of\_Battat (a Shia cleric who opposed them); media organizations \#Furqan\_Foundation (2) and \#Itisam\_Foundation (12); media releases as in \#despite\_the\_disbelievers (3) and \#spreading\_his\_CV (referring to the leader of ISIS) (8); spreading news of battles and areas under their control as in \#Sinai\_State (6), \#Alkhair\_State (9), \#Ein\_Islam (ISIS's name of Kobani) (13), and \#Ramady (15); and the most interesting of them is using popular hashtags, even ones used by their opponents, to spread their message as in \#we\_are\_all\_Moaz (the pilot of the Jordanian plane that ISIS downed) (5) and \#Helal\_Sydney (refers to a soccer match) (14).

The top 15 hashtags (translated from Arabic) used by anti-ISIS tweeps cover: exposure of their violence as in \#ISIS\_stones\_woman (1), \#torturing\_women (10), \#Sydney\_hostages (8), and \#Ahsaa\_shooting (11); using ISIS as a negative label for other groups as in \#MB\_UK (Muslim Brotherhood, United Kingdom) (2); virtual and physical threats against ISIS as in \#banning\_ISIS\_on\_Twitter (5) and \#Egypt\_cemetary\_for\_ISIS (9); using pro-ISIS hashtags to attack the group as in \#KSA\_army\_rule (4) and \#Kurds\_are\_Jewish\_slaves (12); and attacking their sympathizers and leaders as in \#Naef\_Elthaqafi (13) and \#Abu\_Bakr\_Baghdadi (15) respectively.  

Some of the top hashtags used by by pro- and anti-ISIS tweeps seem to be consistent with news stories that appeared on the days when pro- and anti-ISIS tweet-counts peaked respectively. It seems that the top interests of both groups are rather different, except for one incident when an ISIS affiliate group attacked the Egyptian Army in Sinai, leading to the hashtags \#Sinai\_State and \#Egypt\_cemetary\_for\_ISIS for the pro- and anti-ISIS tweeps respectively. There has been discussion in the media about the brutal tactics of ISIS\footnote{\scriptsize \url{http://edition.cnn.com/2014/09/25/opinion/cruickshank-storer-hamid-bakos-isis/}}. Our observations suggest that they seem to galvanize opposition against them, and it is unclear if it is an effective recruitment tool.

\section{Classifying and Analyzing Pro/Anti ISIS Tweeps}
\subsection{Classifying Tweeps}
Given the users we have identified as being most likely supporting or opposing ISIS based on the whether they refer to ISIS (\#\<داعش>) or the Islamic State (\#\<الإسلامية>\_\<الدولة>), we were interested in classifying users as potentially supporting or opposing ISIS \emph{before} they explicitly wrote a tweet identifying their stance.  As mentioned earlier, we automatically labeled 11,332 and 45,628 tweeps as supporting or opposing ISIS.  Of 11,332 supporting tweeps, 4,307 of them indicated support for ISIS starting from the creation dates.  Thus, we could not use them for prediction, and we were left with 7,225 tweeps.  For balance during classification, we randomly picked an equal number of anti-ISIS tweeps whose accounts were active prior to their opposition to ISIS.  We used the tweeps with the tweets they authored prior to the first supporting/opposing tweets for classification.  We randomly selected 10\% of the tweeps for validation, and we performed 10-fold cross-validation on the remaining 90\%.  For classification, we trained a Support Vector Machine (SVM) classifier using the SVM$^{Light}$ implementation with a linear kernel with default parameters.  For features, we used bag-of-words features, including individual terms, hashtags and user mentions.  The Arabic text of the tweets was processed using the method described by Darwish et al.~\shortcite{darwishetal12cikm} to normalize the text and to handle word elongations.  We combined all the tweets for a single user into one document, and we used all the tokens in the combined-tweets documents as features.  We used the validation set to perform SVM threshold adjustment.  The threshold was adjusted to maximize the classification effectiveness (F1 score). 

Table~\ref{classification} lists classification results of potential supporters and opposers of ISIS based on the tweets they authored prior to their first tweet explicitly stating their position.  As results show, both groups are quite separable from each other, even before they voice explicit support or opposition, and the classifier can distinguish between them with relatively high accuracy.

\begin{table}
\begin{center}
{
\begin{tabular}{cccc}
\hline
& Precision & Recall & F1-measure\\\hline
Pro-ISIS & 89.6 & 83.7 & 86.6 \\
Anti-ISIS & 84.7 & 90.3 & 87.4 \\
Average & 87.2 & 87.0 & 87.0 \\\hline
 \end{tabular}
}
\end{center}
\caption{\label{classification}Classification results of potential ISIS supporters/opposers}
\end{table}

\begin{figure*}[ht]
\centering
\includegraphics[width=0.9\textwidth]{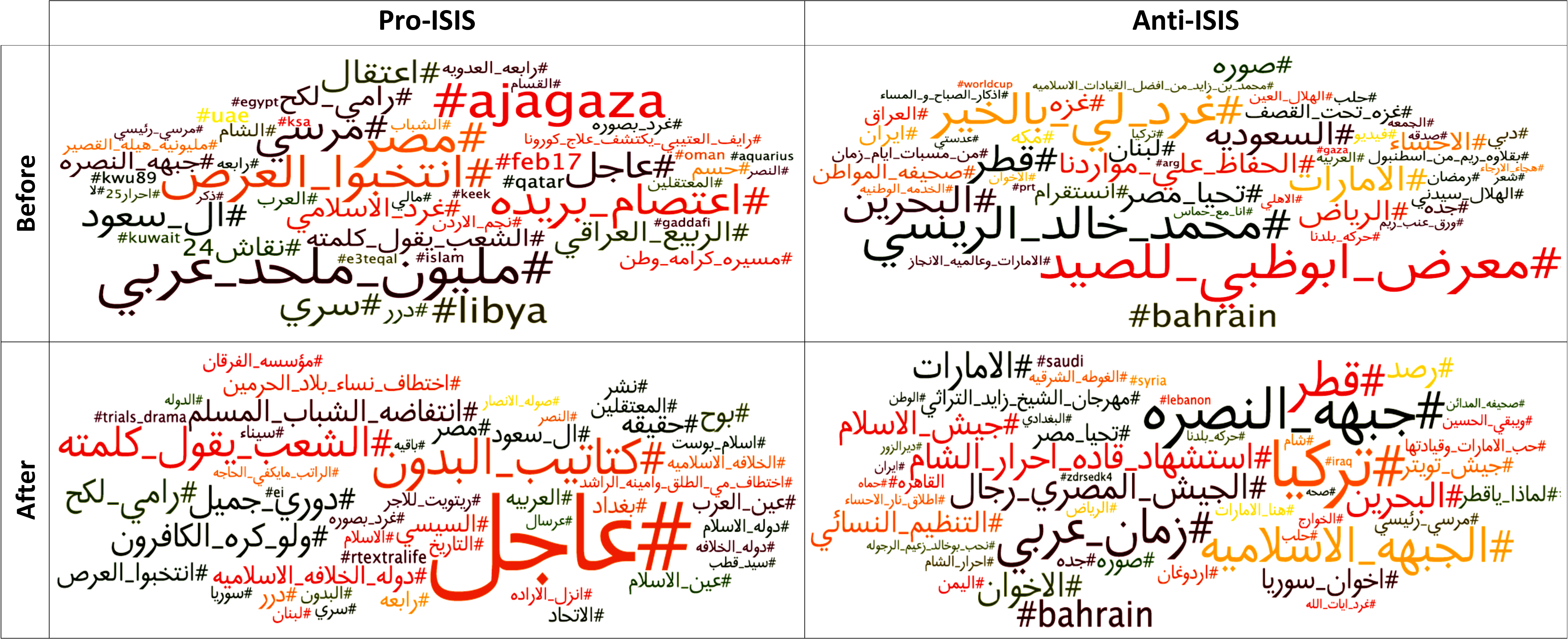}
\caption{\label{Tagclouds}Tagclouds for the most indicative 50 hashtags of ISIS support and opposers in the period before and after they started tweeting about ISIS}
\end{figure*}

Next, we were interested in understanding the underlying features that makes the two groups separable.  So, we consulted the SVM classification model to identify the most distinguishing 50 hashtags that the classifier used to determine if a person would potentially support ISIS and most distinguishing 50 hashtags that would indicate future opposition. We opted to use hashtags instead of individual word for this exploration, because they are more demonstrative of underlying topics.  Figure~\ref{Tagclouds} shows tag clouds for the hashtags for future pro- and anti-ISIS users in the \emph{before} condition.  The size of each hashtag in the cloud is indicative of its weight as assigned by the classifier.

By examining the top 50 hashtags that are most indicative of future ISIS support, we observe the following (each hashtag is followed by its rank on the list):
\begin{itemize}
\item Most hashtags are political and are specifically related to support of the Arab Spring and opposition to regional regimes.  These hashtags by country are as follows (translated to English):\\
\emph{- Egypt}: \#elect\_the\_pimp (4 -- hashtag referring Sisi that emerged during the Egyptian presidential elections), \#Morsi (7 -- ousted Egyptian president), \#Rami\_Lakkah (13 -- Egyptian politician and businessman), \#Rabia\_Adawiyya (29 -- sit-in that was disbanded violently by security forces), \#Rabia (35), \#the\_free\_25 (47 -- referring to Jan. 25 uprising) \\
\emph{- Saudi Arabia}: \#Alburayda\_sit-in (5), \#Saud\_clan (9 -- Saudi rulers), \#the\_people\_say\_their\_word (18), \#finishing\_off (23 -- opposition campaign) \\
\emph{- Iraq}: \#Iraqi\_Spring (12), \\
\emph{- Libya}: \#Libya (6), \#Feb17 (17 -- start date of 2011 revolution), \#Gaddafi (45) \\
\emph{- Syria}: \#Annusra\_Front (19 -- rebel group), \#Levant (25), and \#million\_man\_march\_for\_Qusair (32 -- Syrian city) \\
\emph{- Kuwait}: \#National\_dignity\_march (24 -- protest movement in 2012) \\
\emph{- Jordan}: \#Jordan\_star (31 -- protest movement)
\item Hashtags relating to conflict areas in the Muslim world, namely: \#ajagaza (1 -- Aljazeera Arabic, Gaza), \#Mali (40), and \#Qassam (44 -- Hamas military wing)
\item There are a few hashtags relating to civil-rights or its violation, namely: \#incarceration (11, 43) and \#prisoners (41).  This is in line with previous research that shows that imprisonment and torture seem to ``fan the flames'' of violence~\cite{Bellamy2009}.
\item Only three hashtag is indicative of anything religious namely \#tweet\_Islamic (16), \#Islam (34), and \#remembrance (49)
\item One interesting hashtag was \#a\_million\_atheist\_Arabs that was used for discussions with atheists.
\end{itemize}
In short, most of the hashtags are related to support for the Arab Spring, opposition to existing regimes in the Middle East, solidarity or concern over hotspots in the Muslim worlds, and disillusionment with the current status quo.

As for the top 50 hashtags that are most indicative of future opposition to ISIS, we observe that:
\begin{itemize}
\item Most of the tweets are general like the one about an exhibition in Abu Dhabi (1), \#Instagram (16), \#Reem\_baklava\_from\_Istanbul (29), etc.
\item There are some hashtags expressing political views, namely:\\
- support for the coup in Egypt: \#long\_live\_Egypt (9 -- slogan of pro-Sisi supporters) and \#our\_country\_movement (30 -- a pro-Sisi movement)\\
- support of UAE's political leadership: \#UAE\_international\_performance (32), \#civil\_service (35), \#Mohamed\_bin\_Zayed\_among\_best\_Islamic\_leaders (44)\\
- support of Gaza, which is the only sentiment in common with the other group: \#Gaza (13 and 50), \#Gaza\_under\_attack (20), and \#I\_am\_with\_Hamas (42)\\
- Animosity towards the Muslim Brotherhood: \#Brotherhood
\end{itemize}
As can be seen, most hashtags are not indicative of of any political stances except support for the regimes in UAE and in Egypt.  Members of this group share concern with the future supporters of ISIS in their support for Gaza.

Whereas the previous analysis was done on classifiers built \emph{before} users expressed support or opposition to ISIS, we also built a separate SVM classifier exclusively using data \emph{after} they expressed their opinion. Though this does not directly reveal anything about the antecedents of support, it still helps paint a clearer picture on how the supporters differ from the opposition. For this analysis, tweets containing a reference to ISIS were ignored as, trivially, otherwise the corresponding hashtags would have been the most discriminating features. Figure~\ref{Tagclouds} in the \emph{after} condition shows tag clouds for the most distinguishing hashtags for current pro- and anti-ISIS users respectively. Again, we obtained the hashtags from the SVM classification model.

By examining the top 50 hashtags that are most indicative of current ISIS support, we note the following:
\begin{itemize}
\item Similar to hashtags that they used before declaring their support of ISIS, many of the hashtags express animosity towards different regimes in the Middle East.  Some of these hashtags by country are:\\
\emph{- Kuwait}: \#Bodoon\_schools (2 -- Bodoon are persons without nationality), \#Bodoon (33), \#join\_the\_well (37 -- protest movement) \\
\emph{- Saudi Arabia}: \#the\_people\_say\_their\_word (3), \#Saud\_clan (14), \#kidnapping\_of\_Saudi\_women (20 -- about government incarceration of women), \#trial\_drama (35 -- mocking trials of opposition figures), \#kidnap\_of\_May\_Altalq\_and\_Amina\_Alrashed (42 -- same as \#20), \#salary\_not\_enough (50 -- opposition campaign)
\emph{- Egypt}: \#Rami\_Lakkah (4), \#Muslim\_youth\_uprising (8 -- a day of protest in Egypt), \#elect\_the\_pimp (12), \#Sisi (15), and \#Rabia (21).
\item ISIS specific hashtags including:\\
\emph{- Media releases}: \#Despite\_the\_disbelievers (6) and \#charge\_of\_Ansar (49 -- about attack in Sinai against Egyptian army)\\
\emph{- Other names for ISIS}: \#the\_state (7 and 47), \#Islamic\_caliphate\_state (13), \#Islamic\_caliphate (30), \#State\_of\_Islam (32), \#Caliphate\_state (41)\\
\emph{- Battles related}: \#Ein\_Alarab (18 -- Arabic name for Kobani, a town witnessing ongoing battles), \#Ein\_Islam (22 -- ISIS's name for Kobani) \\
\emph{- ISIS Media organizations}: \#Alfurqn\_Foundation (28), \#Islam\_Post (32)
\item Another carry over hashtag from the \emph{before} condition is \#incarcerated (23).
\end{itemize}

Among the hashtags indicative of opposition to ISIS, there seems to be three main themes, as follows:
\begin{itemize}
\item The first themes includes supportive references to various rebel or Islamic groups in Syria such as \#Alnusra\_Front (2), \#Islamic\_Front (4), \#martyrdom\_of\_Ahrar\_Alsham\_leaders (9), \#Army\_of\_Islam (12), \#Brotherhood\_in\_Syria (15), \#Ahrar\_Alsham (35).  This can be attributed to ISIS attacks on opposition groups in Syria.
\item The second theme includes: \\
1) support for some Middle Eastern regimes, namely United Arab Emirates as in \#Zayed\_cultural\_exhibition (20 -- Zayed is the UAE founder), \#love\_UAE\_and\_its\_leaders (27), \#this\_is\_UAE (28), \#we\_love\_Bu-khaled\_the\_Arab\_leader (44 -- Bu Khaled is vice president of UAE), Egyptian military regime as in \#Egyptian\_army\_are\_men (8), \#long\_live\_Egypt (16), and \#our\_country\_movement (39), and Saudi Arabia as in \#Twitter\_army (18) and \#gunfire\_in\_Ahsaa (36 -- an attack in Eastern Saudi Arabia).\\
2) opposition to various Islamic groups as in \#female\_Organization (14) and \#Brotherhood\_of\_Syria and ISIS as in \#khawarej (37 -- a deviant sect in Islamic history) and \#Baghdadi (42 -- leader of ISIS).\\
3) opposition to regimes that are perceived to support Islamic groups such as \#Zaman\_Arabic (5 -- opposition publication in Turkey) and \#why\_oh\_Qatar (19)
\item The third theme seem to be Shia related as in \#Hussein\_lives\_on (34 -- revered Shia historical figure).
\end{itemize}

\subsection{Example Tweets and Tweeps}
To provide a better qualitative understanding of our data set, we provide some pairs of before/after tweets for ISIS supporters (Table~\ref{beforeAfterExamplesPro}) and ISIS opposition (Table~\ref{beforeAfterExamplesAnti}).

\begin{table*}[ht]
\begin{tabular}{|c|p{2.1cm}|p{14cm}|}
\hline
Tweep & Date & Tweet (translated) \\ \hline \hline
\multirow{3}{*}{T1} & \cellcolor{Gray} May 25, 2012 & \cellcolor{Gray} Don't be surprised if it rains today ... martyrs are spitting on us \\
 & Nov. 9, 2014 & Preliminary schizophrenia: I like ISIS, but I want to watch Chris Nolan's new movie \\
 & Nov. 17, 2014 & The gazes of Bashar's soldiers before slaughter by \#Islamic\_State in \#despite\_the\_disbelivers \\
\hline \hline
\multirow{2}{*}{T2} & \cellcolor{Gray} Aug. 26, 2013 & \cellcolor{Gray} Best pics from the liberation of Khanaser (Syrian town) \#Syria \#Alnusra\_Front \#Ahrar\_Alsham \\
 & Nov. 14, 2014 & The strike of the Islamic State's lions, Sinai State \\ \hline \hline
\multirow{2}{*}{T3} & \cellcolor{Gray} April 9, 2014 & \cellcolor{Gray} \#mb\_europe \#elect\_the\_pimp Some of the military coup crimes in \#Egypt (link) \\
& Nov. 17, 2014 & Praise be to Allah, support for \#Islamic\_State in Indonesian mosques \\ \hline \hline
\multirow{2}{*}{T4} & \cellcolor{Gray} Feb. 24, 2011 & \cellcolor{Gray} JUST IN: Benghazi cleaning the streets! (link) \#Libya \#GaddafiCrimes \#feb17 \\
& Sept. 25, 2014 & Pictures of the dead of Crusader Arab alliance against Islamic State (link) \\ \hline \hline
\multirow{2}{*}{T5} & \cellcolor{Gray} March 9, 2013 & \cellcolor{Gray} An important message from your brother in Syria \#jihad \#Alnusra\_Front \#Free\_Syrian\_Army (link) \\
& Feb. 19, 2014 & Islamic State retakes Babila (Syrian town) after \#Free\_Syrian\_Army betrayal \\ \hline
\end{tabular}
\caption{\label{beforeAfterExamplesPro}Before (light gray) and after (white) tweets from tweeps who end up supporting ISIS}
\vspace{10pt}
\begin{tabular}{|c|p{2.1cm}|p{14cm}|}
\hline
Tweep & Date & Tweet (translated) \\ \hline \hline
\multirow{2}{*}{T6} & Nov. 12, 2014 & Aftermath of Crusader-Arab alliance in Homs countryside \#Alnusra\_Front \#Homs \\
 & Nov. 15, 2014 & Baghdadi state media \#ISIS not different from Bashar state media, all in deceitful swamp \\
\hline \hline
\multirow{2}{*}{T7} & Sept. 9, 2014 & Saddened for loss of heroes \# martyrdom\_of\_Ahrar\_Alsham\_leaders. Same George Bush strategy: long dirty war\\
 & Oct. 30, 2014 & ISIS an internal Muslim problem, to be fixed by Muslim hands, not Americans or their Arab puppets \\ \hline \hline
\multirow{2}{*}{T8} &  Aug. 17, 2014 & Alnusra Front: pictures of your brothers in \#resilient\_Murek (Syrian city) (link) \\
& Aug. 23, 2014 & Disgusting ISIS show to sell their pathetic goods (link) \\ \hline \hline
\multirow{2}{*}{T9}  & Sept.1 2014 & Hilary Clinton agrees with Muslim Brotherhood to establish ISIS \#arrest\_ISIS\_Saudi\_popular\_demand \\
 & Oct. 14 2014 & Why the dead children ! Muslim Brotherhood and ISIS 2 sides of 1 coin \#criminal\_brotherhood \#long\_live\_Egypt \#mb\_uk \\ \hline \hline
\multirow{2}{*}{T10}  & Oct. 24 2014 & Don't let ISIS distract you from worship.  This is there goal. \#Hussein\_lives\_on \\
 & Oct. 27 2014 & From the battlefield against ISIS \#Hussein\_lives\_on (pic) \\ \hline 
\end{tabular}
\caption{\label{beforeAfterExamplesAnti}Tweets from anti-ISIS tweeps ``after'' showing opposition to ISIS}
\end{table*}

Among the ISIS supporters, tweep T1 seemingly supported the 2011 Egyptian revolution and is probably not an Islamist. Tweep T3 opposed the military coup in Egypt. Tweeps T2 and T5 supported different militant groups in Syria and ended up supporting ISIS. In the case of tweep T5, he transitioned from supporting to opposing the Free Syrian Army.  Tweep T4 was a supporter of the 2011 Libyan revolution.

Among the ISIS opposition, tweeps T6, T7, and T8 supported different Muslim rebel groups in Syria and ended up opposing ISIS. Tweep T9 supports the current Egyptian regime and opposes the Muslim Brotherhood.  Tweep T10 is seemingly a Shia fighting against ISIS.

\subsection{Geographic Distribution of Tweeps}
Apart from what users tweets, their geographic location is also expected to be related to whether they are more likely to support or oppose ISIS. To test whether there is indeed geographic variation in patterns of support and opposition we used the users' self-declared locations to map them to countries. For the users labeled as anti-ISIS, 46\% had provided a non-empty location string in their profile. This fraction was only 21\% for those labeled as pro-ISIS. We then matched the user declared locations to a dataset containing country ID's for the 10,000 most commonly used Arabic user locations from a large set of tweets~\cite{mubarak2014geo}. If a location is ambiguous, ex. refers to multiple countries such as ``UK and Syria'', then it was ignored. To further increase the recall of mapped locations, we also used Yahoo Placemaker\footnote{\scriptsize \url{https://developer.yahoo.com/yql/console/#}, then \it{geo.placemaker}.}. Manual inspection showed that this tool worked well for strings in European languages, but it performed poorly for most Arabic strings. Hence, we decided to only pass strings with letters in the Latin alphabet to Yahoo Placemaker. As for annotation, strings where multiple countries were detected were ignored. Using this procedure, we managed to map 52\% of non-empty locations for anti-ISIS tweeps and 21\% for pro-ISIS tweeps. This means that 24\% of the anti-ISIS tweeps could be mapped to a country, compared to 4.4\% for pro-ISIS tweeps. The number of country-labeled pro-ISIS tweeps is rather small, so any further statistics on them are likely inconclusive.  As one might expect, tweeps in the pro-ISIS group are less likely to put real non-empty locations, potentially due to fear of being prosecuted.  Their declared locations are generally nondescript such as ``in the wide world''. Table~\ref{geography} shows the count of countries that we were able to map tweeps to.

\begin{table}[ht]
\begin{center}
\begin{tabular}{|p{3.5cm}|p{3.5cm}|}
\hline
 Pro-ISIS & Anti-ISIS\\\hline
 Saudi Arabia (144) & Saudi Arabia (4,675) \\
 Iraq (56) & Kuwait (1,064) \\ 
 Syria (49) & Egypt (1,060) \\
 Egypt (39) & UAE (753) \\
 Kuwait (15) & Iraq (474) \\
 Qatar (13) & Bahrain (429) \\
 Yemen (12) & United Kingdom (392) \\
 Jordan (12) & Lebanon (321) \\
 Libya (10)  & Qatar (255) \\
 Algeria (9) & Syria (183) \\\hline
 \end{tabular}
\end{center}
\caption{\label{geography}Geographic distribution of support and opposition, derived from self-declared Twitter location strings. Only the top 10 locations for both groups are shown.}
\end{table}

Though large countries such as Saudi Arabia, Egypt and Iraq appear prominently in both lists, there are still noteworthy differences in the relative ranking. Looking at the top five countries for the (\# pro-ISIS/\# anti-ISIS) ratio, one obtains Syria (0.27), Yemen (0.20), Libya (0.13), Iraq (0.11) and Jordan (0.10). The higher the ratio, the higher the relative support for ISIS.  Except for Jordan, all of these are countries are suffering from on-going military conflicts. The two biggest countries in terms of mapped users, Saudi Arabia and Egypt, are both fairly far down the list with (0.03) and (0.04). Only countries with at least 10 mapped pro-ISIS users were used for this analysis. Still, the analysis is ultimately based on a fairly small number of pro-ISIS users and should be more viewed as a proof-of-concept analysis.

\subsection{Focus and Time Analysis of Tweeps}
\begin{figure}
\centering
\includegraphics[width=\columnwidth]{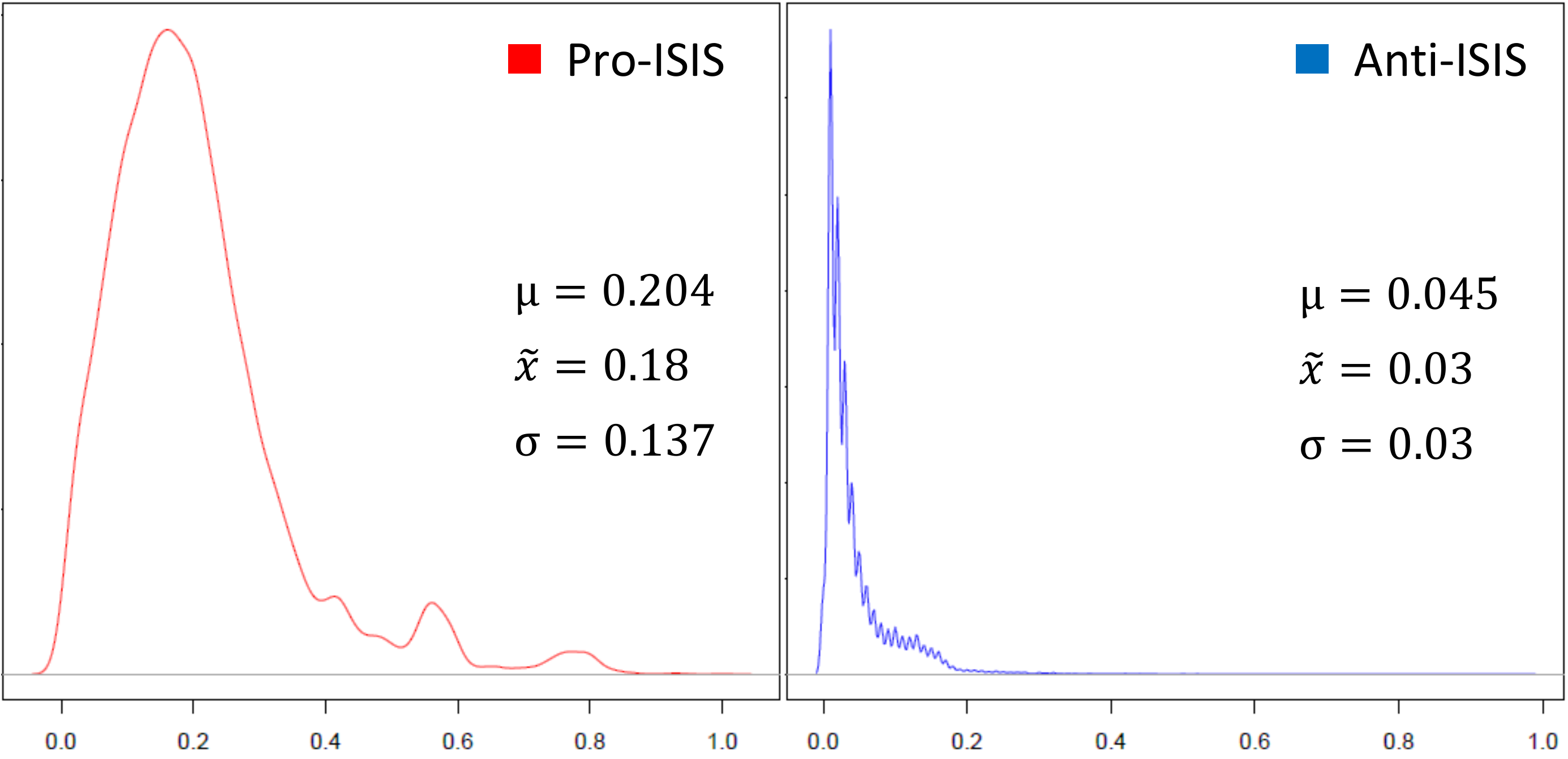}
\caption{\label{Focus}Distribution of users' focus for pro-ISIS vs. anti-ISIS}
\end{figure}

In this section we examine how often pro- and anti-ISIS tweeps tweet about ISIS and the typical creation of their Twitter account.  Figure~\ref{Focus} shows the distribution of ISIS related tweets for both groups.  As can be observed, though the number of pro-ISIS tweeps is significantly smaller than anti-ISIS tweeps, pro-ISIS tweeps dedicate a much large portion of the their tweeting activity to ISIS. On average pro-ISIS tweeps tweet about ISIS 20\% of the time compared to 4.5\% of the time for anti-ISIS tweeps.

\begin{figure}
\centering
\includegraphics[width=\columnwidth]{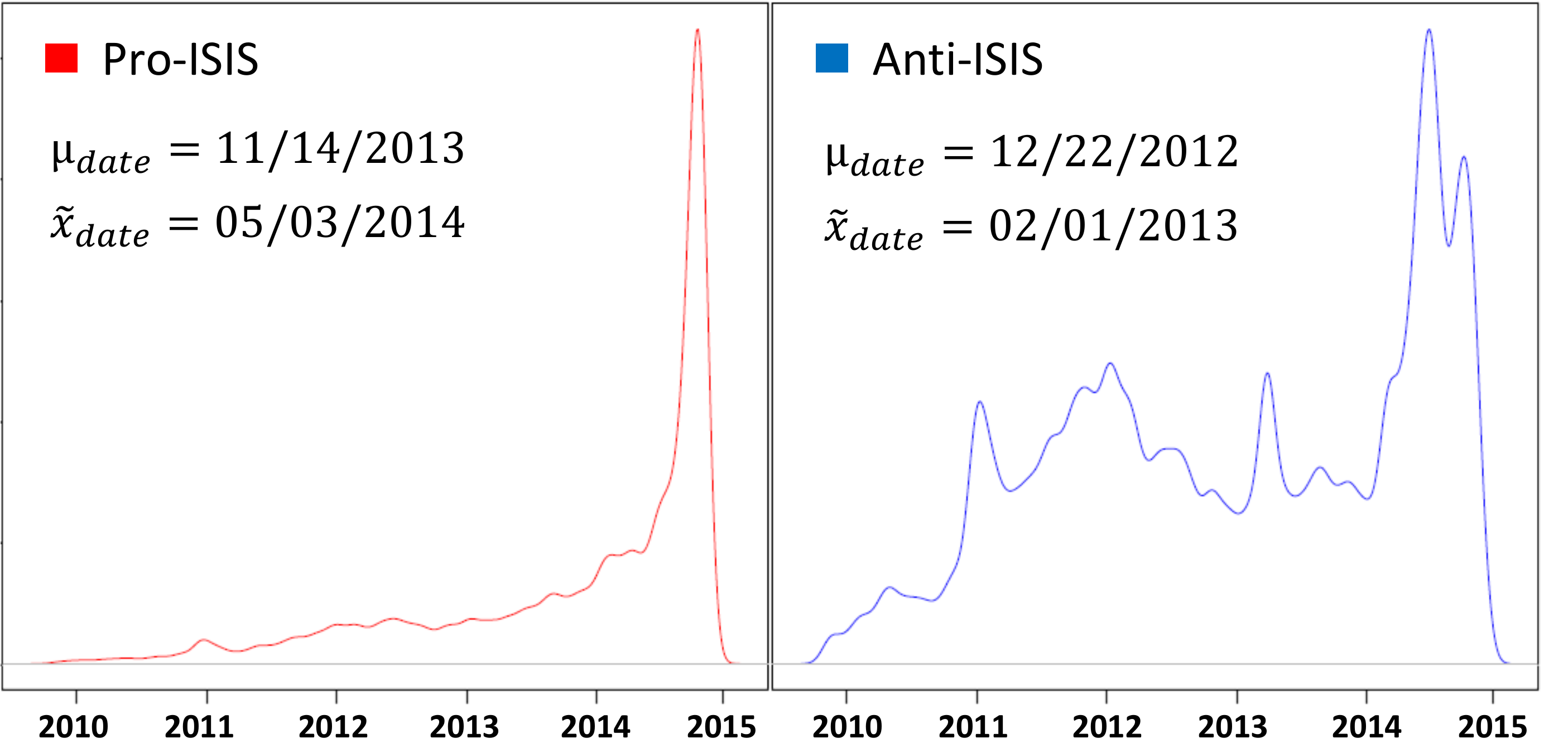}
\caption{\label{Dates}Distribution of users' joining date on Twitter for pro-ISIS vs. anti-ISIS}
\end{figure}

As for when the Twitter accounts were created, Figure~\ref{Dates} provides the distribution of creation dates for both groups.  Concerning the anti-ISIS group, there are two major spikes. The first is in late 2011/early 2012, which is probably caused by the Arab Spring.  The second one that appears in mid 2014 looks interesting and we cannot readily explain it.  More investigation of this is required.  For pro-ISIS tweeps, nearly 32\% of them joined Twitter in September and October 2014.  This spike coincides with the commencement of the US bombing campaign against ISIS.  However, other factors maybe at play and this requires further investigation.

\section{Limitations}
Though our study is closely related to studies understanding the psychology of militants and their supporters, our focus in this paper is on \emph{ideological supporters}, not the actual fighters on the ground. In a sense, we are looking more at the ``soil'' on which violent behavior can flourish. We believe that such a broader view is crucial to finding solutions that go beyond the identification and elimination of individuals.

One of the strengths of our study is that we look at tweets in Arabic, whereas social media studies in general tend to be English-only and US-centric. Most ISIS members are after all Arabs.  However, this also means that our studies does not include any recruiting attempts or press-release-type messages that are deliberately aimed at audiences in the West. Studying such marketing efforts would be interesting in its own right, but deviates from the motivation of our study to understand the antecedents of ISIS support.

Our data collection involved using both the Twitter Streaming API and, after identifying tweets and users of interest, the Twitter REST API to obtain historic information. However, due to the temporal difference of when these two data sets where collected, some accounts had been deleted by the user, had their privacy settings changed to ``protected'' or were suspended by Twitter by the time we collected the historic data.  There has been media reports that suggest that Twitter frequently suspends ISIS related accounts\cite{shane2014isis}. Concretely we failed to get the data for nearly 15 thousand out of 180 thousand tweeps, meaning that our data set is probably biased towards less offensive and less openly hateful users. Despite this shortcoming, our classification results shows that even this set of ``slightly more politically correct'' users could be easily told apart.  Slightly less dramatic than the deletion or suspension of complete accounts, users themselves can delete individual tweets. So if an ISIS supporter was now embarrassed by posting about Britney Spears several years ago, he could go back and delete these tweets. Though a theoretical possibility, we do not expect this kind of behavior to be frequent. A more frequent limitation is the result of Twitter's restriction which only provides up to the most recent 3,200 tweets for a user. Hence, for more active users we could not go back in their history as much as we would have liked.

Needless to say that Twitter users are not necessarily representative of the general population. But especially in these domains surveys would also suffer from non-response bias as many people might prefer not to openly voice their support for a militant organization. In that sense Twitter with its perceived anonymity lends itself to study such phenomena. The fact that our findings concerning the frustration with failed Arab Spring revolutions are in line with previous research~\cite{kohlmann2014profiles,PlousZ05} also suggests that our results are not mere artifacts of a user selection bias.

The representativeness of our study could potentially be improved by broadening the user base by \emph{inferring} a users likelihood to support or oppose ISIS. Classification could use a mix of content and network features and, unlike our current classifier, be applied to users for which the ground truth is not known and might never express their opinion on the topic. Of course, attempting to read the minds of the ``silent majority'' is notoriously difficult \cite{CohenR13}.

Our definition of the ``before'' period could also be seen as a limitation. As the cut-off date between the ``before'' and ``after'' period is different for each user and depends on the date of their first on-topic tweet, it is hard to generalize to the population level. This also relates to the possibility that the temporal dynamics are likely to change over time and that the early supporters might differ from the supporters joining later. The fact that our classifier achieves such good performance suggests though that there are not too many temporal subgroups with vastly different characteristics, though this could be investigated in future work.

\section{Conclusion}
In this paper, we used Twitter data to study the antecedents of ISIS support. At a global, temporal level, we observed some connection between spikes in support or opposition and reports of territorial gains/losses or accounts of torture, forced female enslavement, and similar news. The clearest trend, however, was that ISIS supporters seemed to join Twitter just for the purpose of voicing their support, as could be seen both from temporal concentration, as well as their topical focus on the matter.
At a personal, historic level, we managed to predict future support or opposition of ISIS with 87\% accuracy, using only tweets from the pre-ISIS period. Looking at discriminating hashtags suggested that a  major source of support for ISIS stems from frustration with the missteps of the Arab Spring.  As for opposition to ISIS, it is linked with support for other rebel groups, mostly in Syria, that have been targeted by ISIS, support for existing Middle Eastern regimes, and Shia sectarianism.  We also showed some of the interesting geographical and temporal trends for both pro- and anti-ISIS tweeps.

\bibliographystyle{aaai}
\bibliography{ISISonTwitter}

\begin{thebibliography}{}

\bibitem[\protect\citeauthoryear{Abdulla}{2007}]{Abdulla07}
Abdulla, R.~A.
\newblock 2007.
\newblock Islam, jihad, and terrorism in post-9/11 arabic discussion boards.
\newblock {\em J. Computer-Mediated Communication} 12(3):1063--1081.

\bibitem[\protect\citeauthoryear{Borge-Holthoefer \bgroup et al\mbox.\egroup
  }{2015}]{Borge-HolthoeferMDW15}
Borge-Holthoefer, J.; Magdy, W.; Darwish, K.; and Weber, I.
\newblock 2015.
\newblock Content and network dynamics behind egyptian political polarization
  on twitter.
\newblock In {\em CSCW},  to appear.

\bibitem[\protect\citeauthoryear{Cohen{-}Almagor}{2012}]{Cohen-Almagor12}
Cohen{-}Almagor, R.
\newblock 2012.
\newblock In internet's way: Radical, terrorist islamists on the free highway.
\newblock {\em {IJCWT}} 2(3):39--58.

\bibitem[\protect\citeauthoryear{Louis}{2009}]{Louis2009}
Louis, W.
\newblock 2009.
\newblock If they're not crazy, then what? the implications of social
  psychological approaches to terrorism for conflict management.
\newblock In Stritzke, W. G.~K.; Lewandowsky, S.; Denemark, D.; Clare, J.; and
  Morgan, F., eds., {\em Terrorism and Torture: An Interdisciplinary
  Perspective}. Cambridge University Press.
\newblock  125--153.

\bibitem[\protect\citeauthoryear{Nashmi \bgroup et al\mbox.\egroup
  }{2010}]{AlNashmiCMM10}
Nashmi, E.~A.; Cleary, J.; Molleda, J.-C.; and McAdams, M.
\newblock 2010.
\newblock Internet political discussions in the arab world: A look at online
  forums from kuwait, saudi arabia, egypt and jordan.
\newblock {\em International Communication Gazette} 72(8):719--738.

\bibitem[\protect\citeauthoryear{Plous and Zimbardo}{2005}]{PlousZ05}
Plous, S.~L., and Zimbardo, P.~G.
\newblock 2005.
\newblock How social science can reduce terrorism.
\newblock {\em The General Psychologist} 40(1):1--2.

\bibitem[\protect\citeauthoryear{Sarangi and Alison}{2005}]{Sarangi2005}
Sarangi, S., and Alison, L.
\newblock 2005.
\newblock Life story accounts of left wing terrorists in india.
\newblock {\em Journal of Investigative Psychology and Offender Profiling}
  2:69--86.

\bibitem[\protect\citeauthoryear{Weber, Garimella, and
  Batayneh}{2013}]{WeberGB13}
Weber, I.; Garimella, V. R.~K.; and Batayneh, A.
\newblock 2013.
\newblock Secular vs. islamist polarization in egypt on twitter.
\newblock In {\em ASONAM},  290--297.

\end{thebibliography}

\begin{thebibliography}{}

\bibitem[\protect\citeauthoryear{Abdulla}{2007}]{Abdulla07}
Abdulla, R.~A.
\newblock 2007.
\newblock Islam, jihad, and terrorism in post-9/11 arabic discussion boards.
\newblock {\em J. Computer-Mediated Communication} 12(3):1063--1081.

\bibitem[\protect\citeauthoryear{Adamic and Glance}{2005}]{Adamic2005}
Adamic, L.~A., and Glance, N.
\newblock 2005.
\newblock The political blogosphere and the 2004 u.s. election: Divided they
  blog.
\newblock In {\em LinkKDD},  36--43.

\bibitem[\protect\citeauthoryear{Bellamy}{2009}]{Bellamy2009}
Bellamy, A.~J.
\newblock 2009.
\newblock Torture, terrorism, and the moral prohibition on killing
  non-combatants.
\newblock In Stritzke, W. G.~K.; Lewandowsky, S.; Denemark, D.; Clare, J.; and
  Morgan, F., eds., {\em Terrorism and Torture: An Interdisciplinary
  Perspective}. Cambridge University Press.
\newblock  18--43.

\bibitem[\protect\citeauthoryear{Borge-Holthoefer \bgroup et al\mbox.\egroup
  }{2015}]{Borge-HolthoeferMDW15}
Borge-Holthoefer, J.; Magdy, W.; Darwish, K.; and Weber, I.
\newblock 2015.
\newblock Content and network dynamics behind egyptian political polarization
  on twitter.
\newblock In {\em CSCW},  to appear.

\bibitem[\protect\citeauthoryear{Cohen{-}Almagor}{2012}]{Cohen-Almagor12}
Cohen{-}Almagor, R.
\newblock 2012.
\newblock In internet's way: Radical, terrorist islamists on the free highway.
\newblock {\em {IJCWT}} 2(3):39--58.

\bibitem[\protect\citeauthoryear{Cohen and Ruths}{2013}]{CohenR13}
Cohen, R., and Ruths, D.
\newblock 2013.
\newblock Classifying political orientation on twitter: It's not easy!
\newblock In {\em ICWSM}.

\bibitem[\protect\citeauthoryear{Conover \bgroup et al\mbox.\egroup
  }{2011}]{ConoverRFGMF11}
Conover, M.; Ratkiewicz, J.; Francisco, M.~R.; Gon{\c{c}}alves, B.; Menczer,
  F.; and Flammini, A.
\newblock 2011.
\newblock Political polarization on twitter.
\newblock In {\em ICWSM}.

\bibitem[\protect\citeauthoryear{Darwish, Magdy, and
  Mourad}{2012}]{darwishetal12cikm}
Darwish, K.; Magdy, W.; and Mourad, A.
\newblock 2012.
\newblock Language processing for arabic microblog retrieval.
\newblock In {\em CIKM},  2427--2430.

\bibitem[\protect\citeauthoryear{Horgan}{2003}]{horgan2003search}
Horgan, J.
\newblock 2003.
\newblock The search for the terrorist.
\newblock {\em Terrorists, victims and society: Psychological perspectives on
  terrorism and its consequences}  3--27.

\bibitem[\protect\citeauthoryear{Jamal \bgroup et al\mbox.\egroup
  }{2014}]{jamal2014anti}
Jamal, A.; Keohane, R.~O.; Romney, D.; and Tingley, D.
\newblock 2014.
\newblock Anti-americanism or anti-interventionism: Evidence from the arabic
  twitter universe.

\bibitem[\protect\citeauthoryear{Kohlmann and
  Alkhouri}{2014}]{kohlmann2014profiles}
Kohlmann, E., and Alkhouri, L.
\newblock 2014.
\newblock Profiles of foreign fighters in syria and iraq.
\newblock {\em CTC Sentinel, September} 29.

\bibitem[\protect\citeauthoryear{Liu and Weber}{2014}]{ZheIngmar14}
Liu, Z., and Weber, I.
\newblock 2014.
\newblock Is twitter a public sphere for online conflicts? {A}
  cross-ideological and cross-hierarchical look.
\newblock In {\em SocInfo},  336--347.

\bibitem[\protect\citeauthoryear{Louis}{2009}]{Louis2009}
Louis, W.
\newblock 2009.
\newblock If they're not crazy, then what? the implications of social
  psychological approaches to terrorism for conflict management.
\newblock In Stritzke, W. G.~K.; Lewandowsky, S.; Denemark, D.; Clare, J.; and
  Morgan, F., eds., {\em Terrorism and Torture: An Interdisciplinary
  Perspective}. Cambridge University Press.
\newblock  125--153.

\bibitem[\protect\citeauthoryear{Miller}{2006}]{miller2006terrorist}
Miller, L.
\newblock 2006.
\newblock The terrorist mind i. a psychological and political analysis.
\newblock {\em International journal of offender therapy and comparative
  criminology} 50(2):121--138.

\bibitem[\protect\citeauthoryear{Morstatter \bgroup et al\mbox.\egroup
  }{2013}]{MorstatterPLC13}
Morstatter, F.; Pfeffer, J.; Liu, H.; and Carley, K.~M.
\newblock 2013.
\newblock Is the sample good enough? comparing data from twitter's streaming
  {API} with twitter's firehose.
\newblock In {\em ICWSM}.

\bibitem[\protect\citeauthoryear{Mubarak and Darwish}{2014}]{mubarak2014geo}
Mubarak, H., and Darwish, K.
\newblock 2014.
\newblock Using twitter to collect a multi-dialectal corpus of arabic.
\newblock {\em The EMNLP 2014Workshop on Arabic Natural Language Processing}
  1--7.

\bibitem[\protect\citeauthoryear{O'Callaghan \bgroup et al\mbox.\egroup
  }{2014}]{OCallaghanPGCCC14}
O'Callaghan, D.; Prucha, N.; Greene, D.; Conway, M.; Carthy, J.; and
  Cunningham, P.
\newblock 2014.
\newblock Online social media in the syria conflict: Encompassing the extremes
  and the in-betweens.
\newblock In {\em ASONAM},  409--416.

\bibitem[\protect\citeauthoryear{Plous and Zimbardo}{2005}]{PlousZ05}
Plous, S.~L., and Zimbardo, P.~G.
\newblock 2005.
\newblock How social science can reduce terrorism.
\newblock {\em The General Psychologist} 40(1):1--2.

\bibitem[\protect\citeauthoryear{Poole and Rosenthal}{1985}]{poole1985spatial}
Poole, K.~T., and Rosenthal, H.
\newblock 1985.
\newblock A spatial model for legislative roll call analysis.
\newblock {\em American Journal of Political Science}  357--384.

\bibitem[\protect\citeauthoryear{Sarangi and Alison}{2005}]{Sarangi2005}
Sarangi, S., and Alison, L.
\newblock 2005.
\newblock Life story accounts of left wing terrorists in india.
\newblock {\em Journal of Investigative Psychology and Offender Profiling}
  2:69--86.

\bibitem[\protect\citeauthoryear{Shane and Hubbard}{2014}]{shane2014isis}
Shane, S., and Hubbard, B.
\newblock 2014.
\newblock Isis displaying a deft command of varied media.
\newblock {\em New York Times} 31.

\bibitem[\protect\citeauthoryear{Tyshchuk \bgroup et al\mbox.\egroup
  }{2014}]{TyshchukWLJK14}
Tyshchuk, Y.; Wallace, W.~A.; Li, H.; Ji, H.; and Kase, S.~E.
\newblock 2014.
\newblock The nature of communications and emerging communities on twitter
  following the 2013 syria sarin gas attacks.
\newblock In {\em JISIC},  41--47.

\bibitem[\protect\citeauthoryear{Weber, Garimella, and
  Batayneh}{2013}]{WeberGB13}
Weber, I.; Garimella, V. R.~K.; and Batayneh, A.
\newblock 2013.
\newblock Secular vs. islamist polarization in egypt on twitter.
\newblock In {\em ASONAM},  290--297.

\end{thebibliography}

\end{document}